This manuscript has been authored by UT-Battelle, LLC under Contract No. DE-AC05-00OR22725 with the U.S. Department of Energy. The United States Government retains and the publisher, by accepting the article for publication, acknowledges that the United States Government retains a non-exclusive, paid-up, irrevocable, world-wide license to publish or reproduce the published form of this manuscript, or allow others to do so, for United States Government purposes. The Department of Energy will provide public access to these results of federally sponsored research in accordance with the DOE Public Access Plan.

# Local Superconductivity in Vanadium Iron Arsenide

A. S. Sefat,[1] G. D. Nguyen,[2] D. S. Parker,[1] M. M. Fu,[2] Q. Zou,[2] An-Ping Li,[2] H. B. Cao,[3] L. D. Sanjeewa,[1] L. Li[1], Z. Gai[2]

[1] *Materials Science & Technology Division, Oak Ridge National Laboratory, Oak Ridge, TN 37831, USA*
[2] *Center for Nanophase Materials Sciences, Oak Ridge National Laboratory, Oak Ridge, TN 37831, USA*
[3] *Neutron Scattering Division, Oak Ridge National Laboratory, Oak Ridge, TN 37831, USA*

We investigate the chemical substitution of group 5 into $BaFe_2As_2$ ('122') iron arsenide, in the effort to understand why Fe-site hole doping of this compound (e.g., using group 5 or 6) does not yield bulk superconductivity. We find an increase in c-lattice parameter of the $BaFe_2As_2$ with the substitution of V, Nb, or Ta; the reduction in c predicts the lack of bulk superconductivity [1] that is confirmed here through transport and magnetization results. However, our spectroscopy measurements find a coexistence of antiferromagnetic and local superconducting nanoscale regions in V-122, observed for the first time in a transition-metal hole-doped iron arsenide. In $BaFe_2As_2$, there is a complex connection between local parameters such as composition and lattice strain, average lattice details, and the emergence of bulk quantum states such as superconductivity and magnetism.

High-temperature superconductivity (HTS) in the iron-based materials and cuprates continues to puzzle the scientific community as its ultimate origin remains unknown despite great efforts. This is unlike the sulfur hydrides and related materials, which are rather plausibly considered to be electron-phonon mediated superconductors. In the unconventional HTS, the magnetic order either coexists or is near in composition. BaFe$_2$As$_2$ ('122') is an iron arsenide parent [2] with a room-temperature ThCr$_2$Si$_2$-type tetragonal crystal structure. It is itinerant and only weakly correlated with a spin-density-wave (SDW) antiferromagnetic order below Néel temperature of $T_N$=133 K, closely coupled to an orthorhombic structural transition ($T_N$=$T_s$). The properties of 122 can be altered by in-plane chemical substitutions of Fe using other transition metals such as T=3$d$ (Cr, Co, Ni, Cu), 4$d$ (Mo, Rh, Pd, Ag), or 5$d$ (Ir, Pt, Au) in Ba(Fe$_{1-x}$T$_x$)$_2$As$_2$ [3-12]. Previously, it was found that 3$d$ and 4$d$ dopants belonging to a group in the periodic table produce essentially overlapping temperature-composition ($T$-x) phase diagrams, emphasizing the importance of electron count [3,7,8] over lattice size effects (i.e., 3$d$ vs. 5$d$). Also, it was reported that $d$ 'electron' dopants (e.g., Co [5]) on Fe-site yield superconductivity in 122, while 'hole' dopants (e.g., Cr, Mo [3,4]) do not. Such results are not well understood, especially since very small doping levels of less than 5% (x<0.05) electron dopant can initiate superconductivity. The lack of bulk superconductivity has been described by first principles density-functional theory (DFT) calculations (at x≥0.25 hole dopants in 122) by a modest covalency of $d$ with arsenic $p$ states [4], and hole dopant can induce electronic scattering [10]. Moreover, a recent review of structural and properties demonstrated that $c$ lattice parameter must shrink upon T doping in Ba(Fe$_{1-x}$T$_x$)$_2$As$_2$, if one is to observe 'bulk' superconductivity [1]. In this work, we substitute 122 with group 5 elements of V (<15%), Nb and Ta (~1%), to investigate their structure-property relations. Our hypothesis is that these elements are less likely to carry a large moment (compared to the other hole dopants of Mn and Cr) in Ba(Fe$_{1-x}$T$_x$)$_2$As$_2$ due to their smaller number of unpaired electrons (e.g., V$^{3+}$, $d^2$), and the materials may become superconductors. While we do not find bulk superconductivity, resembling all other reports on hole-doped Ba(Fe$_{1-x}$T$_x$)$_2$As$_2$, we find evidence of local superconductivity for the first time.

Single crystals of Ba(Fe$_{1-x}$T$_x$)$_2$As$_2$ (T=V, Nb, Ta) were grown out of FeAs self-flux technique similar to ref. [13], with [001] direction perpendicular to the crystalline plate shapes. The chemical composition of the crystals was measured with a Hitachi S3400 scanning electron microscope energy-dispersive X-ray spectroscopy (EDS). The structures were identified as tetragonal ThCr$_2$Si$_2$ type (*I*4/*mmm*, *Z*=2) at room temperature, and lattice parameters upon doping were refined using X'Pert HighScore by collecting data on an X'Pert PRO MPD X-ray powder diffractometer (See **Table S1** in **S**upplementary section). **Figure 1** plots *a*- and *c*-lattice parameters as a function of transition-metal concentration in Ba(Fe$_{1-x}$T$_x$)$_2$As$_2$. With V doping, the *a*-lattice parameter decreases while the *c*-lattice parameter increases. In comparison, in the hole doped Cr-122 both *a* and *c* increase, while in Mo-122 the *a*-lattice barely changes while *c* expands [3,4]; for x=0.05, the overall cell volume increases by 0.3% for T=Cr and 0.5% for Mo [3,4], while it only expands by ~0.1% for V. A review of lattice parameter trends in iron arsenides noted that a decrease in *c*-lattice is required to induce 'in-plane' superconductivity [1], and it seems that all in-plane hole-doped (Cr, Mo, V, Nb, Ta, etc.) 122s expand their *c* lattice and should be non-superconductors.

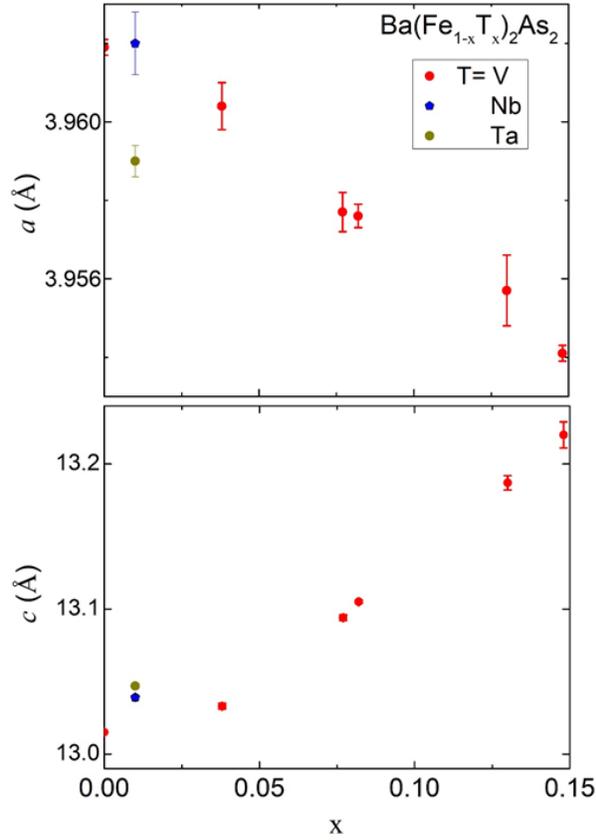

**FIG. 1:** Room-temperature tetragonal $a$- and $c$-lattice parameters for a range of V chemical substitutions (x) below 15%, and for Nb and Ta at 1% level, in Ba(Fe$_{1-x}$T$_x$)$_2$As$_2$.

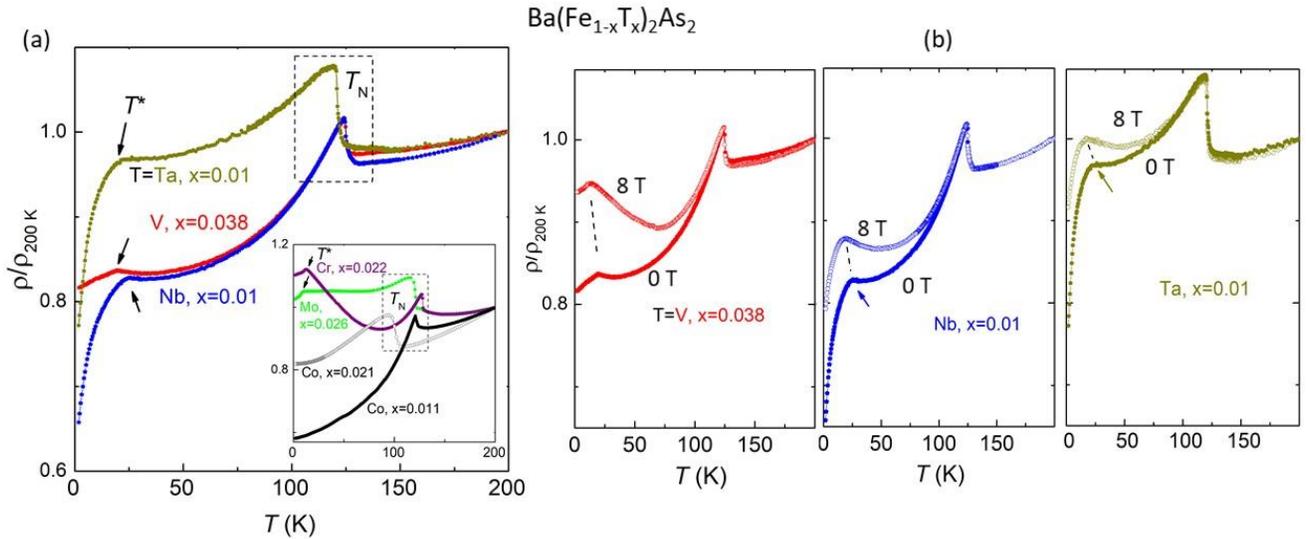

**FIG. 2:** Temperature-dependent resistivity ρ normalized to 200 K data for Ba(Fe$_{1-x}$T$_x$)$_2$As$_2$, T= V, Nb, and Ta. (a) Data at zero field showing anomalies for Néel ordering temperature ($T_N$) and an additional $T^*$ feature that does not reach zero; inset shows $T^*$ that is also seen for other hole dopants of Mo and Cr at small x values, but no such feature for electron dopant of Co. (b) Comparison of data at 0 and 8 T showing field dependence of $T^*$ anomaly for T=V, Nb, and Ta.

Electrical resistance data for Ba(Fe$_{1-x}$T$_x$)$_2$As$_2$ with T= V, Nb, and Ta were collected using Quantum Design's physical property measurement system. The electrical resistance measurements were performed in the *ab*-plane of the crystals, by attaching platinum leads using Dupont 4929 silver paste. The values of resistivity are on the order of a few mΩ cm, here normalized in the form of $\rho/\rho_{200K}$ in **Figure 2**. For each Ba(Fe$_{1-x}$T$_x$)$_2$As$_2$, ρ diminishes from room temperature, with an upturn associated with Néel ordering temperature reflecting the loss of carriers as a partial SDW gap opens. Cooling each sample further, there is a ρ downturn below $T^*$ probably due to the increase in the mobility of the remaining carriers at this temperature. Although $T^*$ was evident in all Nb- and Ta-doped crystals that were measured, it did not appear in all V-doped crystals measured. For Ba(Fe$_{1-x}$T$_x$)$_2$As$_2$, the inferred transition temperatures in ρ for V doping (x=0.038) are $T_N$ =128 K and $T^*$= 19 K, for Nb doping (x=0.01) are $T_N$ =124 K and $T^*$= 24 K, and for Ta doping (x=0.01) are $T_N$ =120 K and $T^*$= 22 K. Hall effect measurements indicate strong hole doping effect through vanadium, even at x=0.016 [14]. The temperature of $T^*$ downturns is reduced upon application of an 8 Tesla magnetic field (**Fig. 2b**), for V-122 giving $T^*$ =13 K, suggestive of a superconducting state. Regarding this mysterious downturn in resistance, the literature shows other lightly hole-doped Cr [4] and Mo [3] 122 crystals having such ρ drops below $T^*$ without reaching zero, while electron doped Co 122 crystals lack this feature (**Fig. 2a**, inset). Such a $T^*$ transition was found to be weakly x dependent, with no structural or magnetic transitions detected via neutron scattering and was argued to result from nanoscale chemical phase segregation or perhaps an impurity inclusion [4]. Since this feature is evident in numerous lightly T hole-doped 122 crystals, we hypothesize that it instead has an intrinsic origin and investigate the local electronic structure for a hole-doped 122 crystal, for the first time here.

To correlate the nanoscale structure with the bulk electronic behavior observed above, we performed scanning tunneling microscopy/spectroscopy (STM/S) measurement on Ba(Fe$_{1-x}$V$_x$)$_2$As$_2$ with x=0.038. The atomic resolution STM image confirms the typical **2×1** surface reconstruction (see supplementary **Fig. S2**). The electronic structure of crystal was examined using current imaging tunneling spectroscopy (CITS) [15]. **Figure 3a** presents atomically resolved STM topographic of an area. The majority of d*I*/d*V* spectra on this area show no-gap like feature as seen in **Figure 3b**. In these spectra, there is a strong nanoscale variation of spectra in this low bias range of −20 mV to +20 mV that might be caused by the local defects in the materials. In addition, we were able to find some local superconducting areas in which d*I*/d*V* spectra show superconducting gap like feature. The typical d*I*/d*V* line spectra across these areas is shown in **Figure 3c** (black arrow). The red curves show a superconducting-like gap with the requisite coherence peaks. In a few unit cells away, the d*I*/d*V* spectra show pseudo-gap characteristics (green curves) and then exhibit normal metallic phase behavior (black curves). These pseudogap states (green curves) are likely the mixture between superconducting and other states. **Figure 3d** shows the temperature evolution of the d*I*/d*V* spectra collected on the same superconducting phase location (blue dot in **Fig. 3a**). With increasing temperature, this gap is suppressed and eventually disappears near 20 K. This 20 K gap suppression temperature is consistent with the downward transition temperature from our bulk resistivity measurements (**Fig. 2a**), suggesting a common origin for these apparently distinct features. **Fig. 3e** plots the temperature dependence of the superconducting gap with a fitted BCS type temperature dependence [16,17]. The gap values with error bars are averaged from seven d*I*/d*V* spectra data set with their standard diviation at various superconducting positions at each temperature.

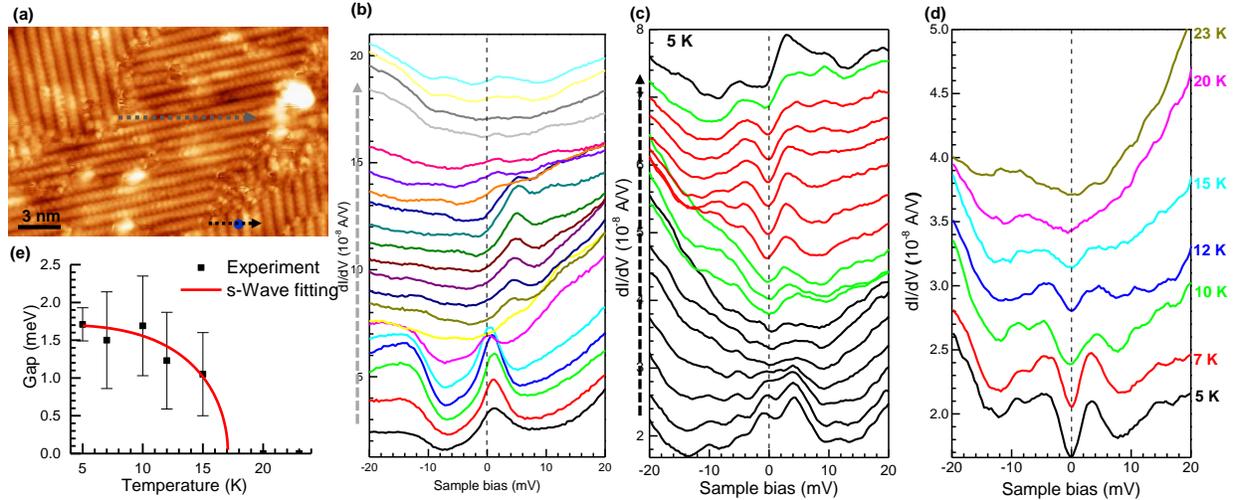

**FIG. 3**. (a) STM topographic image on the surface with **2 × 1** surface reconstruction ($V_S = -20$ mV, $I_t = 200$ pA). (b) d$I$/d$V$ line spectra taken across a regular area marked by the gray dashed arrow in (a); (c) d$I$/d$V$ line spectra taken across a superconducting phase area marked by the black dashed arrow in (a); spatial distance between each curve is 0.21 nm (red and green curves label the superconducting and pseudogap phase spectra, respectively). (d) Temperature dependent d$I$/d$V$ spectra taken at the same location of the blue dot in (a); vertical offsets are applied to the spectra for clarity. (e) Temperature dependence of superconducting gap with s-wave fittings. The gap values was averaged from 7 d$I$/d$V$ spectra data set at various superconducting positions at each temperature. A Dynes model is applied to extract the superconducting gap from d$I$/d$V$ spectra.

The superconducting gap map (**Fig. 4a**) at 5 K is calculated from the CITS, similar to those reported for in-plane electron-doped iron arsenides [15]. The K-means clustering analysis was performed in PYTHON language using PYCROSCOPY package; K-means clustering is a vector quantization data mining approach, which groups a large dataset into components with quintessential characteristics. The clustering method separate the areas studied into superconducting and non-superconducting areas (zero-gap). Although the majority of the crystal shows non-superconducting (zero gap), we detect small localized superconducting regions in the range of one unit-cell up to a few nanometers. The superconducting area are found mostly near domain boundaries and some bright spots on the surface which are due to sub-layer defects as seen in over layer image in **Figure 4b**. In addition, we found some surface areas of Ba(Fe$_{1-x}$V$_x$)$_2$As$_2$ crystal that show non-superconducting gap like feature in d$I$/d$V$ spectra. **Figures 5a, b** show an STM topographic image and d$I$/d$V$ tunneling conductance map from the non-superconducting area. The Fast Fourier transform (FFT) of the conductance map reveals the $q$-space structure through quasiparticle interference (QPI) patterns induced by the scattering of electronic states with defects. The QPI disclose the C$_2$ symmetry of the q-space structure, which is similar to QPI in the SDW state found in other iron arsenides [18]. Four white circles in **Figure 5c** mark the wave vectors corresponding to the nearly square atomic lattice. There are two prominent scattering wave vectors q$_1$ and q$_2$ aligned along the Fe-Fe bond direction which is approximately 45° with the orthorhombic $a$- and $b$-axes. The scattering wave vector q$_1 \approx 2\pi/8a_{Fe-Fe}$ ($a_{Fe-Fe}=a_0/\sqrt{2}$) corresponds to a periodicity of ~ 8$a_{Fe-Fe}$ in real space. The wavevector q$_1$ does not show very clearly which might be due to the scattering of the high density of defects on the sample which obscure small wave vectors. Moreover, we observe another scattering wave vector q$_2 \approx 2\pi/4a_{Fe-Fe}$. Based on the theoretical simulations for QPI in iron arsenides [19,20], the wave vector q$_1$ originates from the intrapocket scattering in a center ellipse-like hole pocket while q$_2$ comes from the interpocket scattering between small electron pockets beside it. Hence, non-percolative local superconductivity in the hole-doped V-122 seems to coexist with antiferromagnetism, which is evident here for the first time.

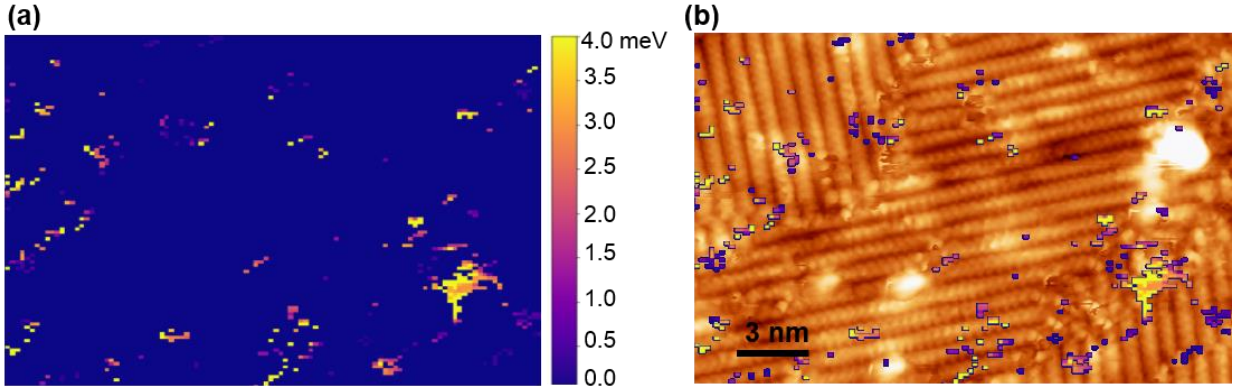

**FIG. 4**. (a) Superconducting gap map extracted from CITS measurement at 5 K. The gap size is calculated from coherent peak position fitting. (b) Overlay image of gap map on top of the STM image as shown in **Fig. 3a**.

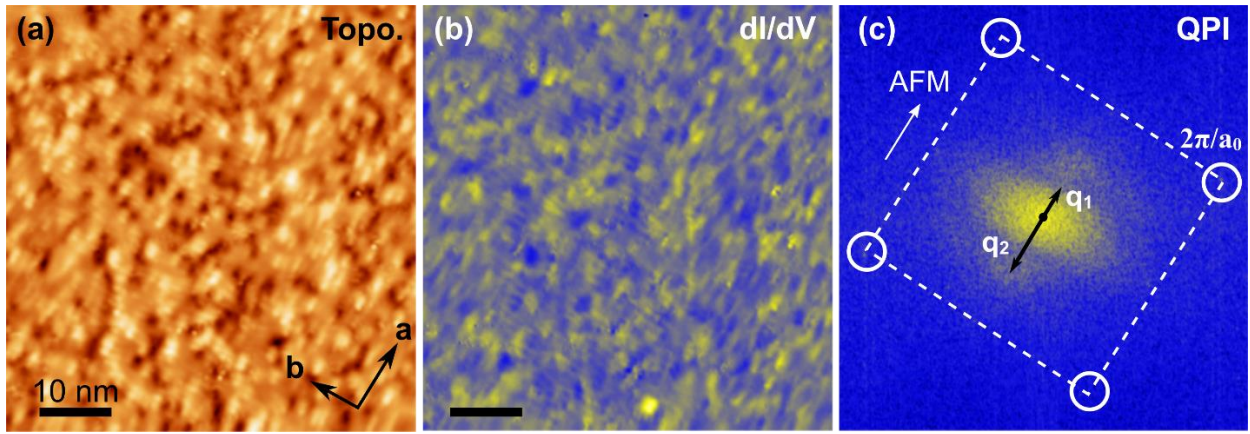

**FIG 5**: (a), (b) STM topographic image and d$I$/d$V$ tunneling conductance map on the surface area with non-superconducting phase ($V_S$ =−5 mV, $I_t$=200 pA, $V_{ac}$=0.5 mV, $f$=973 Hz, $T$=5 K). (c) FFT of conductance map in (b) which demonstrate anisotropic QPI with C$_2$ symmetry.

There are several other pieces of information that taken together comprise a strong case for local superconductivity in this in-plane hole-doped 122 iron arsenide. As shown in the inset of **Fig. 2a**, the resistivity features at $T^*$ drop to a lower temperature under a field of 8 T. While it is unlikely that a spin or charge-density-wave would show such a field-related change in ordering temperature, we can even make a stronger quantitative statement. The Pauli-limited upper critical field $H_{c2}$, in the BCS approximation, for a 20 K superconductor would be approximately 36 T [21]. Assuming linearity in $H_{c2}$, one would then expect an 8 T field to reduce the resistivity feature temperature by approximately 5 K, in general agreement with the 7 K reduction for V-122. Moreover, there is a notable correspondence (most likely not coincidental) between these results and our previous results [5] on electron-doped BaFe$_2$As$_2$: in the work on cobalt-doped BaFe$_2$As$_2$ we observed a superconducting phase with $T_c$ = 21 K for $x_{Co}$= 0.1. Note that, in terms of charge count, this is virtually an equivalent doping (of opposite sign) to our $x_V$=0.037 doping, as V has three fewer valence electrons than Fe, and that the $T_c$ observed there is nearly identical to our $T^*$ here. Furthermore, in that work, we observed under application of the same 8 T field, a very similar reduction in $T_c$. Taken together, all these facts strongly suggest that what we are observing is in fact a local superconductivity.

These local superconducting areas near domain boundaries might share the same origin as in recent proposals in iron arsenide superconductors, in which superconducting states are induced by local strains or interfaces [22-26]. In this case, however, this state is partly induced by vanadium doping. In parallel with the local superconducting phase, the crystal of Ba(Fe$_{1-x}$V$_x$)$_2$As$_2$ with x=0.038 also demonstrates an SDW when scanning at other crystal regions, similar to other iron arsenides [27]. Hence, this non-percolative local superconductivity in V-122 seems to coexist with antiferromagnetism. Given the likely local superconductivity around x = 0.038 for V, it is plausible to argue that there is a nanoscale phase separation associated with the V substitution, with some small regions exhibiting the local superconductivity described previously and the larger fraction maintaining the antiferromagnetic character. In fact, first-principles calculations for vanadium show evidence for stronger scattering compared to cobalt dopants (see **Fig. S3**), which is likely strongest in the physical vicinity of a V dopant atom and weaker at a greater distance. As the doping level is rather low (<4%), this may yield variation in bulk ρ properties in different pieces of as-grown crystals.

We note that this apparent local superconductivity only appears at (and presumably near) the low doping of x=0.038. For samples with larger doping levels there is no hint of the several superconductivity-related features descried above. It is most likely that this is related to the substantial scattering the V dopant atoms produce, as shown in our first principles calculations. In general, superconductivity in 122 materials results from a complex constellation of effects – the role of charge doping in suppressing the magnetic order, changes in unit cell parameters, local strain, the strength of spin fluctuations, and the pair-breaking effects of the scattering due to the dopant atom. It is still an open theoretical question, for example, why doping the usually strongly magnetic Co atom into BaFe$_2$As$_2$ in fact *suppresses* the magnetic order and leads to robust superconductivity. We also note that there is an experimental precedent in BaFe$_2$As$_2$ for this highly doping-dependent superconductivity. Unlike Co and Ni doping, which both produce a strong superconductivity with $T_c$ values exceeding 20 K, Cu doping [28] produces only a minimal ($T_c$ ~ 2 K) superconductivity at the very small doping (quite similar in charge count to x=0.038 of V here) of x=0.044. We have previously argued [9] that this results from strong scattering associated with the Cu atoms, which effectively form a rather distinct band associated with large pair breaking. While the scattering due to V is somewhat weaker than in Cu, the same effect is generally operative here, with the additional distinction that for this small V doping the magnetic state is not fully suppressed. This is most likely due to the unfilled *d*-shell of V, and therefore the magnetism probably interacts with the local superconductivity in some manner. Future theoretical and experimental work is necessary to understand this relationship.

A major unanswered question is the local relationship of the vanadium with the apparent superconductivity. It is tempting to associate the few percent reductions in ρ in the superconducting phase with the few percent (x=0.038) of V present in the crystal and thereby correlate the location of the V atom with the local superconductivity. However, the STM results do not confirm this and in addition, in such a scenario the local environment would be very highly doped, containing an electron count (per cell) of order unity less than the majority phase. This is unlikely to show superconductivity with the $T^*_c$~20 K that we observe. It is more likely that a combination of charge doping, strain, and reduced scattering in the vicinity of the grain boundaries allows local superconductivity to emerge in this crystal, similar to other in-plane hole-doped iron arsenides (T=Cr, Mo, Ta, Nb).

To summarize, we find substantial evidence of novel local superconductivity in vanadium-doped $BaFe_2As_2$ crystals, which appears to coexist, in a phase-separated manner, with antiferromagnetism. Such local superconductivity in an iron arsenide along with coexisting magnetism, is consistent with the lack of bulk superconductivity at higher x for any in-plane hole-dopants and yields additional questions regarding the possibility of a percolative superconducting path in these materials.

**Acknowledgement**

The research is supported by the U.S. Department of Energy (DOE), Office of Science, Basic Energy Sciences (BES), Materials Science and Engineering Division. STM/S study was conducted at the Center for Nanophase Materials Sciences, which is a DOE Office of Science User Facility. This research used resources at the High Flux Isotope Reactor, a DOE Office of Science User Facility operated by the Oak Ridge National Laboratory.